# LA PERVERSIDAD INDUCIDA[1]

Pompeu Casanovas

Profesor de Investigación en IA, Derecho y Ética del Instituto de Investigación en Inteligencia Artificial del Consejo Superior de Investigaciones Científicas (IIIA-CSIC)[2]

Capítulo para el libro *De la ejecución a la historia del Derecho Procesal y de sus protagonistas.* Liber Amicorum *en homenaje al Profesor Manuel-Jesús Cachón Cadenas*. Vol. 5. Carmen Navarro Villanueva, Núria Reynal Querol, Francisco Ramos Romeu, Arantza Libano Beristain, Consuelo Ruiz de la Fuente, Santi Orriols García (Eds.), Barcelona: Ed. Atelier, 2025 (en prensa).

## I. INTRODUCCIÓN

Es un auténtico placer contribuir con este pequeño trabajo al Homenaje al Profesor Manuel Jesús Cachón, colega y amigo desde hace tantos años. Somos de la misma generación, y hemos compartido ideas y trabajos en común, incluyendo la grabación y edición de videos en los tribunales.[3] Nos une además la pasión por la historia intelectual, campo al que ha realizado contribuciones muy importantes. Modestamente, consulté sus trabajos al descubrir el uso del concepto de *juicio*

---

[1] El presente trabajo se enmarca en el Proyecto SGR-532 AQU2021 «Instituto de Derecho y Tecnología». También cuenta como artículo de la Unidad Asociada IIIA-CSIC sobre Derecho e IA.

[2] Asimismo, Profesor Adjunto en la Law School de la Universidad de La Trobe (Melbourne, Australia), e investigador vinculado al Instituto de Derecho y Tecnología de la UAB (IDT-UAB).

[3] Me refiero al proyecto de *Vídeos en los Tribunales de Justicia* (1992-1995) y a *Aula de la Justicia. Video y Guía Pedagógica del Video sobre la Justicia en la Comunidad de Madrid* (2003).

*sumarísimo de urgencia*, aplicado al final de la Guerra Civil española, pero elaborado ya por los juristas de Burgos en 1936 (CASANOVAS Y CASANOVAS, 2009). El Profesor Cachón es también un experto en la historia del derecho procesal catalán. Y me gusta especialmente recordar su reconstrucción de la obra y figura de Josep Antoni Xirau Palau, hermano de Joaquín, uno de los filósofos más interesantes de la denominada "Escuela de Barcelona" en el exilio (CACHÓN CADENAS, 2002, 2011).

En este breve ensayo, sin embargo, voy a centrarme sólo indirectamente en la dimensión histórica. Me referiré en cambio de forma libre y especulativa a las formas sociales y políticas de comportamiento en el mundo inteligente de la web de datos y en el denominado Internet de las cosas.

El título del ensayo, «la perversidad inducida», alude a los efectos perversos de la noción de «compartir» información en las redes sociales cuando ésta resulta falsa —*fake news*— o bien cuando se emiten juicios sobre personas, acontecimientos, empresas o hechos que producen consecuencias negativas que pueden ser consideradas como efectos secundarios, resultados que a menudo conducen a consecuencias que no se previeron inicialmente. Esta contribución complementa el artículo sobre fascismo, agresión e inteligencia artificial (IA) publicado recientemente en el número especial de *TeknoKultura* (UCM) dedicado a las formas de poder político de la inteligencia artificial (IA, en adelante) (CASANOVAS, 2025). En él describí el pensamiento y la ética de Alexander Karp, amigo y socio de Peter Thiel, Doctor en Filosofía por la Universidad Goethe de Frankfurt con una tesis sobre la agresividad, antiguo discípulo de Habermas, y CEO de PALANTIR, la corporación norteamericana de análisis de datos dedicada a la industria bélica. Sugerí implícitamente, sin desarrollarla, la posible conexión entre las formas de gobernanza desarrolladas por las plataformas y la denominada «cultura de la cancelación» [*cancel culture*] o «cultura del llamamiento» [*call-out culture*]. Hay múltiples ejemplos de cómo funciona. Expondré inicialmente el caso de Jonathan Tennant porque no ha sido tratado aún en ningún estudio y puede ilustrar los procedimientos de denuncia en la red; un ejemplo asimismo de conducta agresiva en una comunidad y fuera de ella.

Jon Tennant era un conocido paleontólogo y activista digital defensor de la ciencia en abierto (*Open Science*).[4] Murió de accidente a los 31 años en plena depresión, después de sufrir un severo proceso de cancelación por parte de sus compañeros de comunidad. Tennant era una figura muy significativa, invitado habitual en los congresos de ciencia en abierto organizados por Open Con.[5] Desgraciadamente, en 2015 tuvo un comportamiento inadecuado en una fiesta posterior a uno de ellos y fue denunciado. Intervino el Open Con Code of Conduct (CoC) Committee, quien, después de la investigación, hizo pública su exclusión. Inmediatamente siguió una denuncia anónima más grave de supuesta violación, y una nueva intervención del CoC. Luego llegaron los comentarios, insultos y descalificaciones en Twitter, muchos realizados por desconocidos. Perdió su trabajo, se suspendió la publicación de su libro, y no sólo se cancelaron todas sus conferencias, sino también su imagen pública, su persona. Tuvo el accidente en 2020, cuando intentaba recuperarse de la depresión en Bali. Su hermana Rebecca y algunos de sus amigos empezaron un movimiento de rehabilitación que llegó demasiado tarde. Se le había acusado de delitos graves sin pruebas y, sostenían sus defensores, no había podido responder a las falsedades, ni fue informado de quién le acusaba ni por qué. No había tenido ocasión ni espacio, pues, para alegar su inocencia. Según REBECCA TENNANT (1921a) «de acuerdo con la Ley de Protección contra el Acoso de 1997 (UK), Jon fue víctima de acoso por parte de la multitud de Twitter». Añade una tabla de comparación entre la «justicia en abierto», donde se produce una sistemática inversión de la carga de la prueba, y la justicia procesal-penal, con garantías para el acusado. [6]

---

[4] https://en.wikipedia.org/wiki/Jon_Tennant
[5] Véase la carta abierta encabezada por Tennant y firmada por una veintena de científicos más a la American Association for the Advancement of Science (TENNANT et. al 2015, 2023).
[6] Cf. R. TENNANT (2020a, 2020b, 2021a, 2021b).

No se trata aquí de realizar ningún juicio, ni de tomar partido ni, naturalmente, de criticar la denuncia de comportamientos sexuales impropios, sino de establecer qué es lo que ocurre en un ámbito digital basado en la infraestructura de las plataformas que incentivan a cualquier precio el incremento de los nudos de conexión entre los usuarios y su implicación con ellas. Son las condiciones también para mimetizar el proceso judicial sin garantías produciendo efectos inmediatos que pueden llegar a ser devastadores. Esto es posible porque actuamos a través de la tecnología y, ahora, a través de agentes y sistemas inteligentes que asumimos como extensiones naturales de nuestra cognición. Voy a sugerir que la regulación no se dirija a humanos o a máquinas, sino a su interfaz, a la realidad emergente de la interacción entre humanos y máquinas.

## II. EL PROBLEMA

### II.1. EL CAMBIO EN LAS CONDUCTAS SOCIALES

Hay ciertamente un elemento de indeterminación en la dimensión social y cultural en el que los fenómenos sociales se producen, un componente de imprevisibilidad histórica que es difícil de captar en el momento en que ocurren y son experimentados como vividos. THEODOR ADORNO (1967, 2022, p. 15), en uno de los primeros análisis del nuevo auge de la extrema derecha en las democracias occidentales, ya recogía la observación que en toda democracia surgían figuras políticas extravagantes, conocidas como *lunatic fringe*, en Estados Unidos. Bien, así era considerado Donald Trump a principios del milenio, una figura marginal en un sistema político que creía conocer bien este tipo de fenómenos. Hay preguntas sencillas sin respuesta. ¿Por qué estas figuras marginales han triunfado poco después? Y si han sido democráticamente elegidas, ¿qué comparten los electores con ellos? ¿Y el resto de la población? Hay autores que ya han denominado "política de acoso" al estilo de humillación sistemática impuesto por Trump (TOOZE, 2025). Pero esta forma de violencia no es original, está en los colegios, en los hospitales y —como hemos visto— en las redes sociales desde hace tiempo.

Partiré del supuesto de que no es posible contemplar lo que está sucediendo ahora sin un cierto grado no solamente de consenso sino de intercambio de valores y comportamientos generalizados y socialmente compartidos. Hay que preguntarse por el modo de ser de los ciudadanos y no solamente de los políticos que los representan (o, al menos, reciben sus votos). El desarrollo de Internet y de la web de datos no es ajeno a ello. Hay ya dos generaciones desde su puesta en funcionamiento. La última etapa viene representada por la aplicación de la inteligencia artificial (IA) a sistemas ciberfísicos (robots y robots móviles) y los mismos sistemas de información que contienen grados de autonomía con una capacidad añadida de evolución, de aprendizaje y de transformarse a sí mismos.[7]

Ha habido ya muchísimas advertencias sobre el fin de la civilización basada en los valores de la Ilustración a partir de la sinergia entre los avances de la inteligencia artificial y su uso indiscriminado por parte de la oligarquía tecnológica y, ahora, por parte de una minoría excluyente y exclusiva en el poder. Los economistas también lo han advertido, alarmados por las perspectivas. Desde STIGLIZ (2025) —quien parece confiar en el derecho internacional para evitarlo— hasta la fábula anticipatoria sobre el declive de Estados Unidos debido a la falta de confianza en las instituciones y en su sistema político escrita para *The Financial Times* por el Premio Nobel de Economía de 2024, DAREN ACEMOGLU (2025). La obra de este último es particularmente interesante, puesto que muestra el papel fundamental del Estado de Derecho para el desarrollo económico y social de los países industrializados.

---

[7] Adoptaré aquí la definición de la OECD, recogida también en el reciente Reglamento de IA. «An AI system is a machine-based system that can, for a given set of human-defined explicit or implicit objectives, infers, from the input it receives, how to generate outputs such as makes predictions, content, recommendations, or decisions that can influencing physical real or virtual environments. Different AI systems are designed to operate with varying in their levels of autonomy and adaptiveness after deployment» (OECD, 2024). Cf. Art. 3.1 del Reglamento de IA, UE (2023) «'AI system' means a machine-based system that is designed to operate with varying levels of autonomy and that may exhibit adaptiveness after deployment, and that, for explicit or implicit objectives, infers, from the input it receives, how to generate outputs such as predictions, content, recommendations, or decisions that can influence physical or virtual environments.»

ACEMOGLU (2024) ha podido probar que «si las empresas privadas internalizan sólo una parte de los daños sociales de las tecnologías transformadoras, la adopción del equilibrio tiende a ser demasiado rápida y requiere políticas regulatorias». ¿Qué tipo de políticas? Algunas, como la defensa de la privacidad, la competencia y la responsabilidad son conocidas.[8] Pero, aun así, el modo importa. Es a la ejecución a lo que voy a referirme al final, puesto que, si no se quiere dejar en manos del poder jurisdiccional la aplicación del derecho en circunstancias geopolíticas adversas, hay que pensar otra estrategia de implementación y defensa de los derechos. Voy a sugerir sintéticamente que el cumplimiento (*compliance*) puede ser semiautomatizado o, lo que viene a ser lo mismo, que solamente con inteligencia artificial pueden asumirse los riesgos que la inteligencia artificial comporta. Esto va en contra de la opinión generalizada de que es posible controlar la inteligencia artificial (IA) con los instrumentos jurídicos habituales —regulaciones: leyes, reglamentos y sentencias; e instituciones: parlamentos, gobiernos y judicaturas. No voy a negar que de hecho esto es lo que se debe y se está intentando hacer, pero ¿es suficiente?

El dilema que ya hemos expuesto en otras partes (CASANOVAS y NORIEGA, 2023a, 2023b) es que, por su naturaleza, la IA no puede controlarse solamente desde el exterior, como si sus componentes fueran equivalentes a los ya conocidos de percepción, intelección, memoria y agencia en los seres humanos. Los sistemas de información inteligentes no son humanos, y voy a sostener que *solamente* —no digo exclusivamente— desde la IA puede ejercerse el necesario control y el seguimiento de la IA en tiempo real, dada la complejidad y la no trazabilidad de los procesos algorítmicos —ingesta, memoria, almacenamiento, análisis, razonamiento y creación de nuevos datos.

---

[8] «We highlight two types of policy implications in response to advances in AI technology: policies that affect diffusion patterns and policies that address the consequences of diffusion. The most relevant diffusion- related policy categories are privacy, trade, and liability. Policy design should focus on achieving the desired balance between encouraging diffusion and compromising societal values. As AI diffuses, it will have consequences for jobs, inequality, and competition. Addressing these consequences will be the role of education policy, the social safety net, and antitrust enforcement.» (AGARWAL, GANS Y GOLDFARB, 2023, p. 155).

Los comportamientos y las formas políticas no vienen predeterminadas por la tecnología. La IA puede intensificar los propósitos y materializaciones de cualquier forma política, democrática, oligárquica o dictatorial (LÓPEZ DE MÁNTARAS, 2024). Sin embargo, el diseño de sistemas inteligentes, sean simbólicos, subsimbólicos o una composición entre diferentes tecnologías (*IA compuesta*) favorece una ejecución de funciones mucho más rápida que lo que hace es crear contextos e interacciones nuevas entre humanos y cibersistemas físicos. I.e. crea una sociedad distinta. *Híbrida,* simbiótica, entre los humanos y las máquinas. Y esta relación modifica los patrones de conducta humana tal como antes los habíamos conocido.

## II.2. ACOSO Y CANCELACIÓN

El tema es, pues, la naturaleza humana, y así lo entendió el último DOUGLAS WALTON al plantear el papel de las emociones en la argumentación y los distintos componentes que resultaban necesarios para la reconstrucción de los implícitos presentes en el razonamiento (MACAGNO Y WALTON, 2014; 2018).[9] Suscitar emociones es particularmente importante en política. Términos como 'estúpido', 'perdedor' o 'marxista' pueden ser utilizados y recontextualizados de manera que afecten a las actitudes del interlocutor respecto a un determinado estado de cosas y sugieran un curso de acción (MACAGNO Y WALTON, 2019, p. 230). Es típico de las políticas de acoso, como las de Trump.

Pero en las ciencias del diseño esto puede ir incluso más allá en la construcción digital de los usuarios, puesto que tanto en su conducta real como virtual pueden reproducir estos moldes y convertirse en agentes que los repliquen en un discurso que se retroalimenta y puede llegar a constituir patrones de comportamiento, tanto verbal como cognitivo. En nuestra opinión, ello no solamente afecta a lo que podríamos llamar discursos totalitarios, i.e. discursos *excluyentes* de grupos vulnerables de la población, sino que llegan también a su contrario, a los discursos *incluyentes*, que pueden provocar actitudes intolerantes en la vida cotidiana, en la acción política y en las instituciones. Lo hemos visto al traer a colación el caso de Jonathan Tennant.

---

[9] También VON WRIGHT (1963a, 1963b, 1972), al distinguir el razonamiento *prágmático* (relativo al deíctico personal 'yo', *I*) del *práctico* (en impersonal, *S* o *X*), y TUOMELA (2013) en la separación del modo personal (*I-mode*, *I-thinking*) del colectivo (*We-mode*, *We-thinking*). D'ALTAN, MEYER y WIERINGA (1996) distinguieron entre el razonamiento ideal (*ideal*) y el real (*actual*) mediante la distinción entre enunciados de deber-hacer (*ought-to-do statements*) y enunciados de deber-ser (*ought-to-be statements*). Los primeros expresan imperativos de la forma 'un actor debe realizar una acción'. Los segundos expresan una situación o estado de cosas deseable sin tener que mencionar a los actores y acciones que tienen relación con esta situación.

Hay que ir con cuidado al analizar este fenómeno. Aún no sabemos con certeza de qué modo está conectado con lo anterior. Parece que es un fenómeno que afecta a los consumidores de todo tipo, ya sean empresas, corporaciones o individuos. Lo hemos visto en artistas y estrellas de la televisión que han cometido errores (SCHEINBAUM Y POEHLMAN, 2025). Pero las políticas denominadas de la *cancelación* —o *cultura de la cancelación*—que hoy son aplicadas en las redes y grupos sociales han sido también asumidas como instrumento político normalizado por los activistas que defienden políticas alternativas.[10] Tienen la característica de que se lanzan contra la persona, contra individuos que de un modo u otro se considera que han dado motivos de reprobación y que, muchas veces, no pueden responder porque no saben de dónde viene la acusación y cómo pueden defenderse. Este hábito que empezó siendo proactivo y ha acabado siendo agresivo se ha convertido en rutina moral. Se «comparte» primero la información o juicio negativo sobre alguien. Así se destruye la imagen pública de quien se reprueba, e indefectiblemente el efecto es devastador para la persona. Esto alimenta el espacio público que se ha transformado a partir de Internet y ha llegado a informar el comportamiento institucional. Incluso instituciones educativas de prestigio —como Berkeley, Stanford y Harvard— han asumido estas premisas, «cancelando» a varios de sus profesores de más nivel. Incluso del más alto nivel, como es el caso del filósofo vivo más respetado, John R. Searle (1932-), suspendido de sus privilegios a los 89 años por la Universidad de Berkeley.

---

[10] Sobre sus orígenes y desarrollo, cf. NG (2020), BURGOS Y HERNÁNDEZ DÍAZ (2021), BRODER (2024). Vid. una comparación desde el punto de vista del procedimiento con las garantías penales, en CABRERA PEÑA Y JIMÉNEZ CABARCAS (2021).

## III. DERECHOS Y ECOSISTEMAS JURÍDICOS

### III.1. RAZONES DE LOS CAMBIOS EN LA CONDUCTA

¿Qué es lo que puede explicar este tipo de conductas generalizadas en la red? ¿Y hay algún modo eficaz de evitarlas?

No existe una respuesta general para una pregunta de este tipo. Alude a un fenómeno social complejo cuya respuesta cultural se halla abierta y veremos como evoluciona en el futuro. Notemos enseguida que la conducta no sólo sucede *online*; sucede *offline* también de manera muchas veces simultánea.

Creo que la transformación del modo de relacionarse en sociedad es un signo más que una causa. Hay una serie de variables societales de entre las cuales es difícil escoger la más relevante o significativa. Se ha hecho un gran esfuerzo en psicología social para identificar y describir los problemas de conducta individuales y colectivas producidos tanto por la economía precaria de las plataformas y las aplicaciones web como por la dependencia relacional inducida en las redes sociales por comunicadores y *youtubbers* varios.

Hay una gran variedad de fenómenos que se dan al mismo tiempo pero que no deberían confundirse: (i) los efectos producidos por la precariedad y dureza del trabajo precario y temporal; (ii) los efectos de cansancio (*stress, burnout...*) debidos a la extracción de esfuerzo en la tecnificación estructural de las condiciones de trabajo; (iii) los efectos de adaptación a los requisitos exigidos por el uso de servicios sociales (desde la educación a la sanidad); (iv) los efectos de agresividad en los adolescentes y jóvenes; especialmente en las escuelas, institutos y universidades; (v) el anonimato como facilitador de la expresión de la agresividad.[11]

Los estudios divergen entre sí, con metodologías y formas diferentes de selección de muestras, en distintos países y continentes. Pero a lo que parecen apuntar todos casi de forma unánime es a la dureza del contexto al que deben enfrentarse los individuos. La sociedad digital ha incrementado las formas de extracción de trabajo y, consecuentemente, ha endurecido la presión que reciben los individuos y los grupos sociales. Ha agudizado, en suma, las diferencias de oportunidad, encaje y ascenso social. Y ha obligado a una adaptación masiva de sus formas de acceso y comunicación.

---

[11] Cf. SUN (2023), LIU et al. (2023), DENG et al. (2024), LU et al. (2024), ZHENG et al (2024), MANE et al. (2025), REHMAN (2025).

Esto también vale para los servicios púbicos como hospitales, escuelas y universidades, donde se ha precarizado e incrementado el trabajo de los más jóvenes al asumir la organización el sistema de plataforma. Puede combinar su funcionamiento con los sistemas en la nube: plataformas como servicio (PaaS) e infraestructuras como servicio (IaaS).[12] Puede también adoptar diversos modelos de negocio (RANERUP, HENRIKSEN Y HEDMAN, 2016). Otra de sus características es la inversión de la carga de trabajo, que ahora realizan los usuarios. La sobrecarga y el fenómeno de *burnout* están sobradamente documentados (FERNÁNDEZ-SUÁREZ et al. 2021, PATE et al. 2023).

Resulta pertinente subrayar que la cultura superior no es un indicador fiable para medir la probidad del comportamiento. Sorprende un poco que la «justicia web» sea obra precisamente de universitarios con un alto grado de preparación. Quizás esta sensación de hallarse por encima del derecho arrogándose la capacidad de juzgar forme parte de las «creencias de lujo» recientemente señaladas por ROB K. HENDERSON (2024) como propios del nuevo capital social de la distinción y el simbolismo de las élites egresadas de las universidades que confieren estatus a bajo coste para los implicados. Pero mi intuición es que se trata también de una conducta transversal, desarrollada desde los reducidos nichos de actuación que han dejado libres las plataformas como posibilidad de elección de los usuarios de los servicios. Es, pues, el reverso complementario de las conductas obedientes que buscan las plataformas, la otra cara de la *regimentación* tecnológica que practican habitualmente, como veremos en las próximas secciones.

---

[12] El National Institute of Standardization and Technology (NIST) define IaaS como "the capability provided to the consumer is provision processing, storage, networks, as well as other fundamental computing resources where the consumer is able to deploy and run arbitrary software, which can include operating systems and applications. The consumer does not manage or control the underlying cloud infrastructure but has control over operating systems, storage, and deployed applications; and possibly limited control of select networking components (e.g., host firewalls)."

## III.2. EL CAMBIO EN LA NOCIÓN DE USUARIO

La tesis es que lo que realmente ha cambiado con la economía de plataforma (*platform-driven economy*) y la aplicación de la IA *es la personalidad del usuario*. Esto no es perceptible a simple vista, pero es lo que sucede y, en parte, ha sido previsto, incentivado y fomentado por la economía de las grandes plataformas como Google o Twitter (ahora X). Ello implica no modelar sino moldear sus gustos, inducir sus deseos y, en definitiva, adquirir más control sobre las decisiones que puedan tomar; i.e. modificar sus capacidades de adaptación y disposición cognitiva. Ése es el objetivo de mercado real, lo que un diseñador debe aprender y tener claro si no quiere ver peligrar su lugar de trabajo.

Así, el significado del término 'usuario' ha cambiado desde que se popularizó en los años ochenta del siglo pasado (NORMAN y DRAPER, 1986). Se tenía en cuenta entonces la experiencia de quien 'usaba' el producto, sus 'disposiciones' ante el objeto diseñado —*affordances*, para usar el término de GIBSON (1966). El producto era *para* los usuarios. Ahora sucede al revés: en un mundo plenamente digitalizado los usuarios son *para* el producto. La memoria acumulada de los profesionales (UX, *user experience designers*) permite entenderlo: «ya no diseñamos realmente experiencias para los humanos».[13] En la mayoría de los casos, pese a las tendencias de investigación centradas en *human-centered design* (HCD), las empresas no tienen tiempo para incorporar al usuario final. Éste ya no forma parte del proceso de diseño. La IA generativa ha acelerado esta tendencia y

---

[13] Sobre el cambio de significado de las funciones de la profesión (UX), POWELL (2018) escribe: « The most disappointing thing about this is that in the main we're no longer designing experiences for humans, not really. There are UX roles out there, quite a few of them, where the end user is not part of the process at all. No research, no testing, the UXer is supposed to just know what their user wants, because it's either as complex as magic or alternatively as easy as common sense. At best they're expected to see the falloff point in a Google Analytic data report and know instantly what the problem is. »

ha dotado de nuevos instrumentos a los desarrolladores. El antiguo ingeniero de Google JAMES W. WILLIAMS ha descrito del modo siguiente su experièncía:

«Pronto comprendí que la industria tecnológica no estaba diseñando productos, sino diseñando usuarios. Estos sistemas mágicos y de uso general no eran "herramientas" neutrales, sino sistemas de navegación orientados a un propósito que guiaban vidas humanas de carne y hueso. Eran extensiones de nuestra atención. (…) Los nuevos desafíos que enfrentamos en la Era de la Atención son, tanto a nivel individual como colectivo, desafíos de autorregulación. […] Pero también sabía que no se trataba solo de mí: de mi libertad, mi atención, mis profundas distracciones, mis objetivos frustrados. *Porque cuando la mayoría de las personas en la sociedad usan tu producto, no solo estás diseñando usuarios, estás diseñando la sociedad* [énfasis añadido].»

Creo que tiene razón. En mi opinión, hay un reto ético —colectivo y de encaje de principios en los sistemas de computación— y otro reto moral, básicamente de *auto-gobierno individual*. La IA tiene el reto de facilitar, de ayudar a proporcionar los instrumentos para la conducción interna de nuestro propio yo. Esto es lo que no estamos consiguiendo.

## III.3. EL CAMBIO EN LOS INSTRUMENTOS JURÍDICOS

El cambio de sociedad no va a dejar indiferentes a las categorías y procesos jurídicos. Y tampoco lo que hasta ahora hemos entendido por el trabajo conceptual de los juristas. Es especialmente la concepción moral del derecho —por qué deben obedecerse o seguirse las normas, i.e. de qué modo éstas se insertan en los contextos sociales— lo que se desvanece con rapidez en la economía conducida por plataformas o, lo que es lo mismo, en la *regimentación* tecnológica que éstas inducen.[14] Técnicamente, «regular» implica considerar a los agentes artificiales como autónomos. Pueden decidir si cumplir o violar las normas. «Regimentar», en cambio, implica reducir su autonomía e imbuir el cumplimiento dentro de su ámbito de acción, reduciendo el de decisión. Literalmente, no pueden violar las normas. Esta distinción puede aplicarse asimismo a los agentes humanos. Si quiere o debe usar el servicio, el usuario está obligado a seguir pasos pautados que no puede modificar y deben ser validados por el sistema. Parece que el interfaz, la interacción entre humanos y máquinas, esté regulado, pero lo que sucede en realidad es que se halla, en gran parte, regimentado. Y lo que produce esto a la larga es un cambio en la disposición cognitiva del usuario, que se acostumbra a ello y tiende a reproducir este tipo de requerimientos fuera del entorno digital. ¿Hay algún efecto en el modo de operar de los sistemas jurídicos? Creemos que sí, y de forma ambivalente, puesto que la transformación no tiene por qué ser negativa para la regulación del mundo híbrido que he venido describiendo y que no podemos soslayar.

---

[14] He descrito con más precisión la diferencia entre regulación y regimentación en CASANOVAS (2025).

Por una parte, tal como ya ha notado COLIN RULE, el antiguo director de resolución de disputas en línea (ODR) de eBay and PayPal, «la mayoría de la gente utiliza cada día herramientas tecnológicas para completar sus listas de tareas pendientes, por lo que ahora esperan que también puedan recurrir a ellas para resolver cualquier problema que encuentren» (RULE, 2020, 277). Por otra, los instrumentos del derecho ya han empezado a construirse de forma que incorporen el máximo de escenarios y grados de gobernanza jurídica, es decir, la implementación (o «aplicación») de normas en la multitud de contextos que requieren sistemas de regulación desde la perspectiva de los sujetos de derecho (como trabajadores, consumidores, ciudadanos...).

Por «gobernanza jurídica» entendemos el conjunto de procesos que generan un ecosistema de regulación sostenible capaz de plasmar los conceptos y valores del estado de derecho en tiempo real, incluyendo su dimensión ética (CASANOVAS et al., 2025). Un ecosistema jurídico inteligente es el sistema que se genera a través de procesos de información en plataformas y aplicaciones mediante el uso de tecnologías avanzadas, como las de la IA. En nuestro trabajo reciente, hemos tenido ocasión de construirlos para los procesos de producción dentro de las fábricas de las manufacturas de la Industria 4.0 (denominadas *smart manufactures*) (CASANOVAS, 2024).[15] De hecho, es la única manera de efectuar la conexión normativa práctica entre humanos y máquinas cuando se trata de regular el flujo de información que sigue un ciclo descendente y ascendente desde los sensores y actuadores a los módulos de la infraestructura del sistema, y de ésta hasta la nube. Así converge la cadena de valor de la IA con el valor añadido buscado por la industria.

---

[15] Véase también POBLET et al. (2019), CASANOVAS et al. (2024).

Esto implica traducir en parte las normas jurídicas y los principios éticos desde su expresión originaria en una lengua natural hasta su formalización en reglas ejecutables en un lenguaje de programación. De hecho, hace ya más de quince años que las corporaciones tecnológicas lo hacen para guiar, i.e. regimentar, la conducta de sus usuarios. Hace diez años, el director de CODEX de Stanford, MICHAEL GENESERETH, sugirió que la administración pública debía hacer lo mismo.[16] Originó el movimiento denominado *Rules as Code* —«las normas como programa»—, por el que las máquinas pueden «consumir» reglas jurídicas ejecutables para facilitar los trámites de los ciudadanos. En parte, la sugerencia ha tenido éxito y ha sido adoptada por muchas administraciones (incluidas las de la Unión Europea). Aunque la dificultad estriba en la interpretación normativa para producir código y, otra vez, la defensa de los derechos de los ciudadanos.[17] Parece prudente seguir utilizando el lenguaje natural para expresar las disposiciones legislativas, frente a quienes proponen hacerlo directamente mediante lenguajes de programación.

La pregunta ahora es si estas tecnologías intermedias, junto a otras provenientes de la «conformación» del derecho por parte de sistemas de inteligencia artificial generativa —*Law informs Code* (NAY, 2022)— están en situación de resolver problemas sociales complejos como el que supone la cultura de la cancelación.

Mi intuición es que no lo están, al menos de momento. Pero su desarrollo quizás podría contribuir a pensar las condiciones de control que podrían ser aplicadas a la tecnología utilizada por las grandes corporaciones y plataformas de la comunicación. Este es el problema: hay que actuar dentro de su diseño. Esto implica imbuir valores ejecutables en los sistemas multi-agente que constituyen *instituciones online* con diversos grados de autonomía (como Google, X, Netflix o Meta-i.e. Facebook, entre muchas otras). Así, las prácticas y los medios convencionales pueden adaptarse para la gobernanza de los sistemas de IA a través de dotarles de una autonomía moral alineada con valores éticos y de derechos humanos (NORIEGA et al. 2023). El simple esquema de la Figura 1 puede ayudar a visualizar esta idea.[18]

---

[16] https://www.youtube.com/watch?v=HELWJwuIA3s (última visita 17.2.2025)
[17] Para una visión crítica de *Rules as Code*, CASANOVAS et. al. (2019).

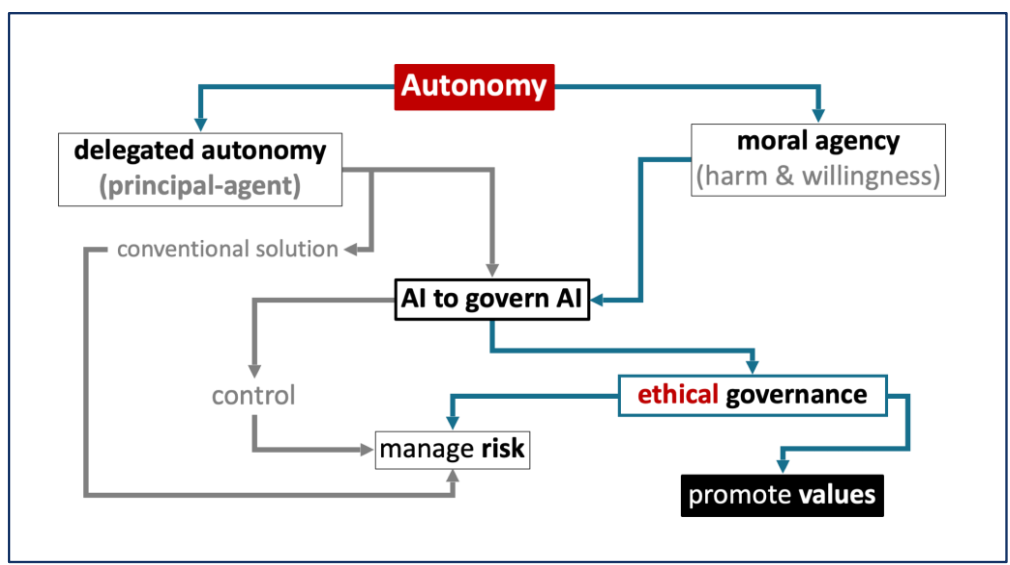

Fig. 1. Esquema para la gobernanza ética de (y a través de) la IA. Fuente: NORIEGA Y CASANOVAS (2024).

¿Constituye esto una solución? Aún no; si bien ya se han presentado diseños de plataformas para la educación y el trabajo en línea que van en esta dirección y sistemas multi-agente con una gran carga ética. Creo que representa un progreso en la forma de concebir la gobernanza mediante IA, pero no puede suplir la voluntad de los usuarios de los servicios ofertados por las corporaciones, ni confrontarse con la falta de voluntad de éstas para diseñar sistemas autorregulados que sean eficaces para su control. Es pues, un problema político antes que jurídico. Depende de quién quiera y pueda ponerlo en práctica. Dudo que sean los gigantes tecnológicos. Podría empezarse más modestamente desde abajo, desde las asociaciones sociales, grupos implicados de derechos humanos, sindicatos e instituciones educativas que hayan tomado conciencia del problema.

---

[18] Es compatible y se puede comparar con los más elaborados en CASANOVAS et al. (2024) que no reproduzco aquí.

## IV. A MODO DE CONCLUSIÓN

Voy a efectuar a modo de conclusión algunas observaciones que me limitaré a apuntar. Se refieren al comportamiento social y político inducido por las plataformas.

Aún más que los potenciales riesgos de la autonomía de los sistemas inteligentes, lo que resulta preocupante es el factor humano, i.e. la forma en que los individuos y grupos humanos cambian a través de su relación con los sistemas inteligentes. Por eso he sugerido que la regulación debe apuntar a las simbiosis entre humanos y máquinas. Una de las mejores formas de hacerlo es la construcción de ecosistemas éticos y jurídicos que permitan la defensa y ejercicio de los derechos, contrarrestando las formas regimentales propias de las plataformas de servicios.

Para ello es necesario aprender a imbuir valores en los sistemas inteligentes. Desde el punto de vista técnico, no es nada sencillo y, entre otras cosas implica la construcción de instituciones en línea (*online institutions*) que actúen en tiempo real. La dificultad básica estriba en el modelado de la ética de la virtud. Dicho brevemente, en el supuesto que pueda hacerse, los valores pueden ser tanto positivos como negativos. Así, un sistema multi-agente puede incorporar tanto la maldad como la bondad, y qué es lo que haga va a depender en parte de su diseño, pero también de propiedades emergentes de su implementación en contextos reales. Nada asegura que su comportamiento corresponda al diseño de las intenciones. Esto, como tantas otras cosas, ya había sido notado por CRISTIANO CASTELFRANCHI (2001) al elaborar la teoría de las sociedades artificiales. Una veintena de años más tarde, el problema de las propiedades sociales emergentes persiste.

La agresividad, pues, si no la propia propensión a la violencia como mecanismo de respuesta y regulación, puede ser uno de los resultados colaterales imprevistos no solamente de la educación, sino de la adaptación al mundo regimentado. Lo primero valdría para los jóvenes de la generación digital. Pero resulta que la conducta agresiva no les es exclusiva. Abarca franjas de edad mucho más altas. Si esto es así, estamos ante la versión contemporánea del antiguo mecanismo social del ostracismo y la lapidación, los mecanismos de los ordenamientos vindicatorios estudiados por antropólogos jurídicos como RAYMOND VERDIER y, entre nosotros, por IGNASI TERRADAS (2008). Es una hipótesis que valdría la pena explorar, aunque la integración comunitaria que subyace a los ordenamientos vindicatorios de las sociedades antiguas y no occidentales no se da de la misma manera. Al contrario, ahora se acepta que el mecanismo de integración es en realidad de distribución de recursos y riesgos, sin atender a los daños colaterales. El derecho vindicatorio, en cambio, es derecho, derecho institucionalmente creado y aceptado por la comunidad. Contrariamente, la cultura de la cancelación se caracteriza en su espontaneidad por la ausencia de derecho. El clic, el *tweet* o el rechazo no responden a una justicia comunitaria, aunque pueden incentivar conductas que se consoliden como pautas de conducta y hábitos no censurados. Es esta ambigüedad entre lo tolerado y lo no censurado lo que valdría la pena explorar. Y esto podría regularse si apuntamos en la buena dirección de lo que cumplimiento, *compliance*, significa: la consolidación del puente entre humanos y máquinas.

## V. BIBLIOGRAFÍA